\documentclass[twocolumn,aps,prl,superscriptaddress]{revtex4}

\usepackage{amsmath}
\usepackage{graphicx}
\usepackage{MnSymbol}

\begin{document}

\title{A simple model for viscoelastic crack propagation}

\author{B.N.J. Persson}
\affiliation{PGI-1, FZ J\"ulich, Germany, EU}
\affiliation{www.MultiscaleConsulting.com}

\begin{abstract}
When a crack propagate in a viscoelastic solid energy dissipation can occur very far from the crack tip
where the stress field may be very different from the $r^{-1/2}$ singular form expected close to the crack tip.
Most theories of crack propagation focus on the near crack-tip region.
Remarkable, here I show that a simple theory which does not account for 
the nature of the stress field in the near crack-tip region
result in a crack propagation energy in semi-quantitative agreement
with a theory based on the stress field in the near crack-tip region.
I consider both opening and closing crack propagation, and show that for closing crack
propagation in viscoelastic solids, some energy dissipation processes must occur in the
crack tip process zone.
\end{abstract}

\maketitle

\setcounter{page}{1}
\pagenumbering{arabic}




\vskip 0.3cm
{\bf 1 Introduction}

The cohesive strength of solids usually depend on crack-like defects, and the energy to propagate cracks
in the material. Similarly, the strength of the adhesive bond between two solids is usually determined by the energy to propagate interfacial
cracks. Here we are interested in crack propagation in viscoelastic materials, such as 
rubber\cite{Knaus2,Gent,Creton,Kramer1,HuiX,Gennes,Brener,Crack1,CP,Green1}.
This topic is of great importance, e.g., for adhesion\cite{Gent}, or for the wear of tires or wiper blades, which result from the removal 
of small rubber particles by crack propagation\cite{wear}.

When a crack propagate in a viscoelastic solid energy dissipation can occur very far from the crack tip
where the stress field may be very different from the $r^{-1/2}$ singular form expected close to the crack tip.
Most theories of crack propagation focus on the near crack-tip region, where the stress field takes the $r^{-1/2}$ singular form.
Here I show that neglecting the detailed form of the stress field close to the crack tip
result in a crack propagation energy in semi-quantitative agreement
with a treatment which includes the singular stress field in the near crack-tip region.

\begin{figure}[tbp]
\includegraphics[width=0.4\textwidth,angle=0]{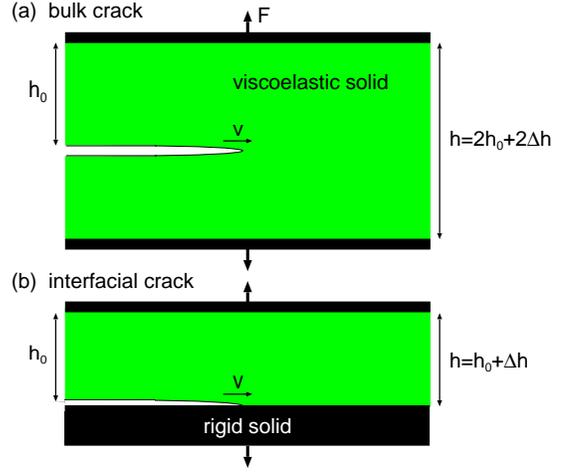}
\caption{
(a) Crack propagation in the bulk of a viscoelastic solid (cohesive crack propagation), and (b) at the interface
between a viscoelastic solid and a countersurface (adhesive crack propagation).
}
\label{crackpicslab.eps}
\end{figure}

\vskip 0.3cm
{\bf 2 Theory of crack propagation in viscoelastic solids}

A crack in a viscoelastic solid can propagate in the bulk or at an interface.
For a bulk crack (see Fig. \ref{crackpicslab.eps}(a)) the stress and strain are usually very high close to the crack tip
and nonlinear effects, involving the breaking of strong covalent bonds, chain pull-out and and cavity formation, will occur close to the
crack tip. This region of space is denoted the crack-tip process zone, the detailed nature of which is an active research field.

Interfacial crack propagation occur in many applications, e.g., between rubber materials and a hard
counter surface as for pressure sensitive adhesives (see Fig. \ref{crackpicslab.eps}(b)). In this case the strain and stresses at a crack 
tip can be much smaller, in particular if the interaction
at the interface is dominated by the weak van der Waals interaction. In this case nonlinear viscoelastic effects 
may occur only in a very small region close to the crack tip where the bond breaking occurs. However,
for very soft materials, like the weakly crosslinked rubber compounds used in pressure
sensitive adhesives, strongly non-linear effects (such as cavitation and stringing)
may occur in a large region close to the crack tip\cite{Leonid,PSA}.

\vskip 0.2cm
{\bf A. Viscoelastic modulus}
 
\begin{figure}[tbp]
\includegraphics[width=0.25\textwidth,angle=0]{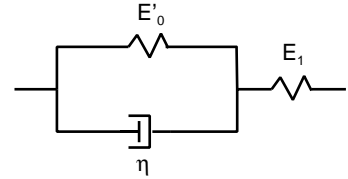}
\caption{
Three-element viscoelastic model
used in model calculation of the crack propagation energy $G(v)$.
The low frequency modulus $E(0)=E_0=E_0'E_1/(E_0'+E_1)$ and the high frequency modulus $E(\infty)=E_1$
and the viscosity $\eta$ are indicated.
}
\label{rheologypic.eps}
\end{figure}

\begin{figure}[tbp]
\includegraphics[width=0.45\textwidth,angle=0]{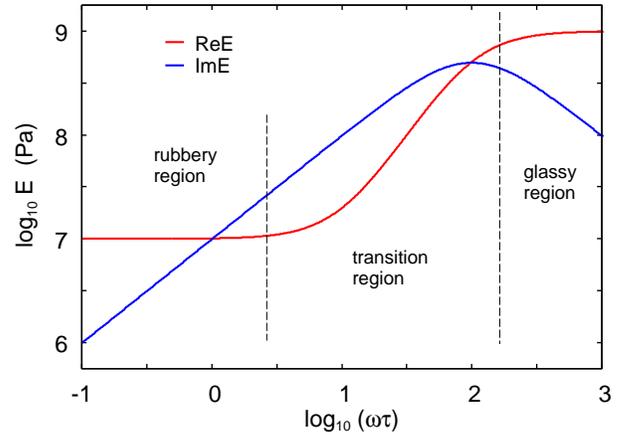}
\caption{
The real and the imaginary part of the viscoelastic modulus as a function of frequency $\omega$ (log-log scale).
For the three-element model shown in Fig. \ref{rheologypic.eps} with $E_1=10^7 \ {\rm Pa}$ and $E_1=10^9 \ {\rm Pa}$.}
\label{1logOmega.2logE.eps}
\end{figure}

Assume that a rectangular block of a linear viscoelastic material is exposed to an
elongation stress $\sigma (t)$. This will result in a strain
$\epsilon(t)$. If we write
$$\sigma(t)=\int_{-\infty}^\infty d\omega \ \sigma(\omega ) e^{-i\omega t}\eqno(1)$$
$$\epsilon(t)=\int_{-\infty}^\infty d\omega \ \epsilon(\omega ) e^{-i\omega t}\eqno(2)$$
then
$$\sigma(\omega )=E(\omega) \epsilon (\omega)\eqno(3)$$
For viscoelastic materials like rubber the viscoelastic modulus $E(\omega)$ is a complex quantity, where the imaginary part is related
to energy dissipation (transfer of mechanical energy into the disordered heat motion). 
In the study below we will use the three-element rheological model illustrated in Fig. \ref{rheologypic.eps}.
For this model the viscoelastic modulus
$$E= {E_0 E_1 (1-i\omega \tau) \over E_1-i\omega \tau E_0}\eqno(4)$$
In Fig. \ref{1logOmega.2logE.eps} we show the dependency of $E(\omega )$ on frequency (log-log scale). 

For low frequencies (or high temperatures) the rubber respond as a
soft elastic body (rubbery region) with a modulus $E(\omega)$ of order $\approx 1 \ {\rm MPa}$ for the rubber used in tires or
$\approx 1 \ {\rm kPa}$ for the weakly cross-linked rubber used in pressure sensitive adhesive films. At very high frequencies
(or low temperatures) is behaves as a stiff elastic solid (glassy region) with the Young's modulus 
$E(\omega)$ of order $\approx 1 \ {\rm GPa}$. In the transition region it exhibit strong internal damping
and this is the region important for energy loss processes, e.g., rubber friction. 
Real rubber exhibit a broad distribution of relaxation times, rather than the single relaxation time as in (4), but already the
simple three-element model exhibit the basic physics of relevance here.

\begin{figure}[tbp]
\includegraphics[width=0.45\textwidth,angle=0]{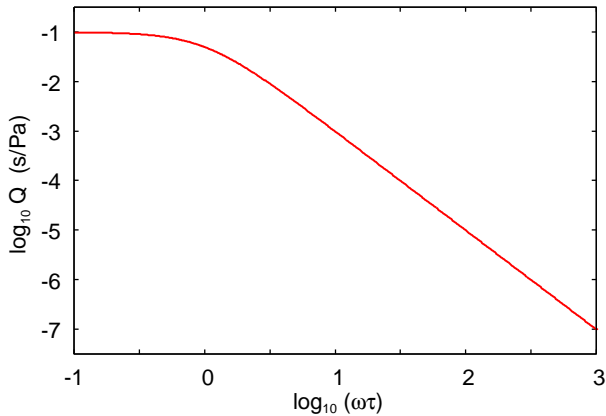}
\caption{
The crack loss-function $Q(\omega)=(1/\omega){\rm Im}[1/E(\omega)]$ as a function of the frequency $\omega$ (log-log scale).
For the three-element model shown in Fig. \ref{rheologypic.eps}.}
\label{1logOmega.2logCrackLossFactor.eps}
\end{figure}

The viscoelastic modulus $E(\omega )$ is a causal linear response function. This imply that the real and the imaginary part of 
$E(\omega )$ are not independent functions but given one of them one can calculate the other one using a Kramers-Kronig equation.
One can also derive sum-rules, and the most important in the present context is
$${1 \over E_0} -{1 \over E_1}= {2\over \pi} \int_{0}^\infty d\omega {1 \over \omega} {\rm Im} {1\over  E( \omega )},\eqno(5)$$
and
$$E_0 - E_1= {2\over \pi} \int_{0}^\infty d\omega {1 \over \omega} {\rm Im} E( \omega ),\eqno(6)$$
where $E_0=E(0)$ is the static ($\omega =0$) modulus, and $E_1=E(\infty)$ the modulus for infinite high frequency $\omega = \infty$.
The function 
$$Q(\omega)={1\over \omega} {\rm Im} {1\over E(\omega)}\eqno(7)$$
occurring in the integral in (5) is very important in viscoelastic crack propagation, and we will denote it as the the crack-loss-function.
It is shown in Fig. \ref{1logOmega.2logCrackLossFactor.eps} for the same model rubber
as in Fig. \ref{1logOmega.2logE.eps}. Note that $Q(\omega )$ decay monotonically with increasing 
frequencies, and is hence largest in the rubbery region in spite of the small magnitude of the damping in this frequency region.

\vskip 0.2cm
{\bf B. Viscoelastic energy dissipation in rectangular strips}

The energy dissipated per unit volume when a strip of material is (dynamically) stretched is given by
$$U =  \int_{-\infty}^\infty dt \ \dot \epsilon (t) \sigma (t)$$
Using (1) and (2) and that
$$\int_{-\infty}^\infty dt \ e^{-i(\omega +\omega') t} = 2 \pi \delta (\omega +\omega')$$
we get
$$U =  2 \pi \int_{-\infty}^\infty d\omega \ (-i \omega)\epsilon (\omega) \sigma (-\omega)\eqno(8)$$
Using $\sigma (\omega ) = E(\omega) \epsilon (\omega)$ we get
$$U =  2 \pi \int_{-\infty}^\infty d\omega \ (-i \omega)\epsilon (\omega) E(-\omega ) \epsilon (-\omega)$$
$$= 4 \pi \int_{0}^\infty d\omega \ \omega |\epsilon (\omega)|^2 [-{\rm Im} E(\omega)]\eqno(9)$$
and using $\epsilon (\omega) = \sigma (\omega )/E(\omega)$ gives
$$U =  2 \pi \int_{-\infty}^\infty d\omega \ (-i \omega){\sigma (\omega) \over E(\omega )}\sigma (-\omega)$$
$$=  4 \pi \int_{0}^\infty d\omega \ \omega |\sigma (\omega)|^2 {\rm Im} {1\over E(\omega)}\eqno(10)$$

\begin{figure}[tbp]
\includegraphics[width=0.45\textwidth,angle=0]{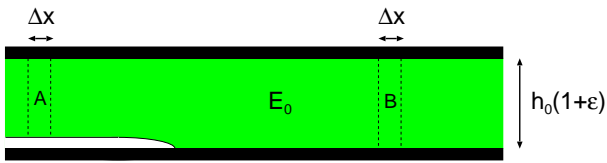}
\caption{
Crack in an elastic solid. For an opening crack the elastic energy stored in the segment B (of width $\Delta x$) 
is used to break the bonds in a surface area
of width $\Delta x$ (transition ${\rm B} \rightarrow {\rm A}$). For a closing crack the gain in 
surface energy when the surfaces close over a region of width $\Delta x$ is used to
stretch the strip A (of width $\Delta x$) (transition ${\rm A} \rightarrow {\rm B}$).
}
\label{newdashedopencloseElastic.eps}
\end{figure}

\begin{figure}[tbp]
\includegraphics[width=0.45\textwidth,angle=0]{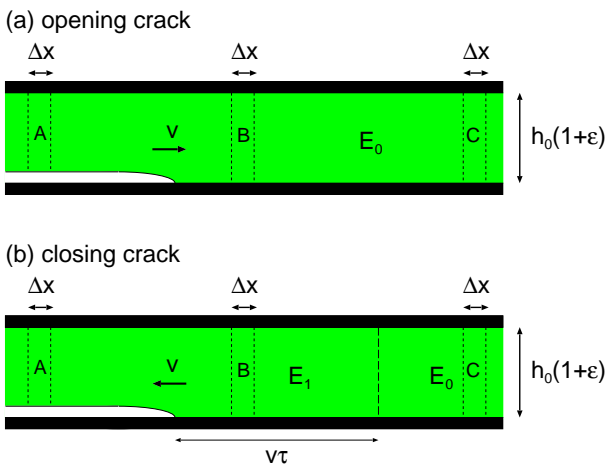}
\caption{
Fast moving opening crack (a) and closing crack (b) in thin viscoelastic slab under tension. In (a) 
the viscoelastic energy
dissipation result in an effective crack propagation energy $G\approx (E_1/E_0)G_0$ which is enhanced by a factor $E_1/E_0$. 
In (b) viscoelastic energy 
dissipation result in an effective crack propagation energy $G\approx (E_0/E_1)G_0$ which 
is reduced by a factor $E_0/E_1$ (see text for details). 
}
\label{OpenClosePic.eps}
\end{figure}

\vskip 0.2cm
{\bf C. Viscoelastic crack: qualitative discussion}

When an opening crack propagate in the bulk of a viscoelastic solid the breaking of the bonds in the crack tip process zone is
usually an irreversible process: the broken (dangling) bonds formed during the crack opening 
react quickly with molecules from the atmosphere, or with mobile molecules
in the solid. Hence if the external crack driving force is removed no closing crack propagation involving the reformation of the original
bonds will occur. However, for interfacial crack propagation the situation may be very different. Thus, in many cases rubber bind to
a countersurface mainly with the weak and long-ranged van der Waals bonds. In this case the bonds broken during  
crack opening and the bonds formed during crack closing may be very similar, and we will assume this to be the case in what follows.

Consider first a crack in an elastic solid. 
We consider the set-up illustrated in Fig. \ref{newdashedopencloseElastic.eps}.
Both sides of a rectangular slab of an elastic material are bonded to rigid plates. The rigid plates are displaced so
that the height of the elastic solid increases from $h_0$ to $h_0+\Delta h= h_0(1+\epsilon_0)$,  where the
strain $\epsilon_0 = \Delta h/h_0$. Assume now that an interfacial crack occur and let $\gamma$ be the energy per unit area to
break the bonds between then the solids at the lower interface. For a stationary crack energy conservation  require that the
elastic energy stored in the strip of width $\Delta x$ far in front of the crack tip is equal to the energy to break
the bonds at the interface i.e.
$$\gamma \Delta x = {1\over 2} \sigma_0 \epsilon_0 h_0 \Delta x = {1\over 2} h_0 E_0 \epsilon_0^2 \Delta x\eqno(11)$$
For an elastic solid, neglecting emission of elastic waves from the crack tip, 
(11) is valid both for stationary and moving (opening or closing) cracks.

Consider now a crack in viscoelastic solid. For a stationary crack the condition (11) is still valid,
where $E_0$ is the static (or low frequency) modulus. However, stationary
cracks are of no real interest as they will not result in failure of the material. For a moving crack in a viscoelastic
solid (11) is no longer valid because of viscoelastic energy dissipation. If $P$ denote the viscoelastic energy dissipation
per unit time, then when the tip has moved the distance $\Delta x = v \Delta t$ the viscoelastic energy dissipation
equal $P\Delta t$. For an opening crack the energy conservation condition becomes 
$$\gamma \Delta x + P\Delta t= {1\over 2} h_0 E_0 \epsilon_0^2 \Delta x$$
or
$$\gamma v + P_{\rm open}= {1\over 2} h_0 E_0 \epsilon_0^2 v$$
For a closing crack we get instead
$$\gamma v = P_{\rm close}+{1\over 2} h_0 E_0 \epsilon_0^2 v$$
Physically, for a closing crack the energy gained by the binding of the solids at the crack interface is 
in part lost as viscoelastic energy dissipation inside the solid. 
The energy to propagate the (opening or closing) crack is given by the elastic energy stored far away from the tip
and is denoted by $G$:
$$G={1\over 2} h_0 E_0 \epsilon_0^2$$
Thus we get
$$G_0 v + P_{\rm open}= G_{\rm open} v$$
$$G_0 v = P_{\rm close}+G_{\rm close} v$$
where $G_0=\gamma$.

For an opening crack, as the crack speed $v\rightarrow \infty$
we have $G/G_0 \rightarrow E_1/E_0$ but for a closing crack $G/G_0 \rightarrow E_0/E_1$. These results 
can be understood by considering the simple crack problem shown in Fig. \ref{OpenClosePic.eps}.

Fig. \ref{OpenClosePic.eps} shows a fast moving opening crack (a) and closing crack (b) in a thin viscoelastic slab under tension. 
In case (a) the slab is elongated by $h_0 \epsilon_0$, 
and we wait until a fully relaxed state is formed before inserting the crack. 
Thus the elastic energy stored in the strip C of width $\Delta x$ is
$\sigma_0 \epsilon_0 h_0 \Delta x/2=E_0 \epsilon_0^2 h_0 \Delta x/2$. 
This energy is partly used to break the interfacial bonds and partly dissipated due to the material viscoelasticity.

Consider a slab of material of width $\Delta x$ as it moves from one side of the crack to the other side.
During this transition it will experience a (elongation) stress $\sigma(t)$ which for a very fast moving crack
can be considered as a step function $\sigma = \sigma_0$ for $t<0$ and $\sigma = 0$ for $t>0$, where $t=0$ correspond to
the case where the segment $\Delta x$ is at the crack tip.

The work done by the external force acting on the segment $\Delta x$ must equal
the energy which is used to break the bonds in the segment of width $\Delta x$. 
If we denote this work with $U h_0 \Delta x$ then
$$U=-\int_{-\infty}^\infty dt \ \sigma \dot \epsilon = -\sigma_0 \int_{-\infty}^0 dt \ \dot \epsilon
= \sigma_0 \left [\epsilon_0-\epsilon (0)\right ]$$
where we have used that $\epsilon=\epsilon_0 = \sigma_0 /E_0$ for $t=-\infty$. Now the strain $\epsilon (0)$ for $t=0$ is
actually undefined because the stress make a step-like change at $t=0$. 
One can show that the correct way to make $\epsilon (0)$ well-defined is to use
$$\epsilon(0) = {1\over 2} \left [\epsilon (0^+)+\epsilon (0^-)\right ]$$
where $0^+$ and $0^-$ are infinitesimal positive and negative numbers. Since $\epsilon (0^-)=\epsilon_0$ and
$$\epsilon(0^+) = \sigma_0\left ({1\over E_0}-{1\over E_1}\right )$$
where we have subtracted the instantaneous reduction in the strain due to the instantaneous (high frequency) elastic response
(with modulus $E_1=E(\infty)$). Thus we get
$$U= {1\over 2} {\sigma_0^2 \over E_0} -{1\over 2} \sigma_0^2 \left ({1\over E_0}-{1\over E_1}\right )$$
For a crack in an elastic solid (neglecting energy dissipation from phonon emission from the crack tip)
$E_0=E_1$ and we get the standard result that the elastic energy $\sigma_0^2/(2 E_0)$ can be fully used to break the bonds
at the crack tip, but in the present case (for a fast moving crack)
$$U={1\over 2} {\sigma_0^2 \over E_1}$$ 
and the condition $U h_0 \Delta x = G_0 \Delta x$ gives
$${1\over 2} h_0 {\sigma_0^2 \over E_0} {E_0\over E_1} = G_0$$
or $G=G_0 E_1/E_0$. 

For the closing crack (case (b)) the
situation is different: For a fast moving crack the strip A is quickly elongated when it approach the crack tip, which require a large stress
$\sigma=E_1 \epsilon$ determined by the high frequency modulus $E_1$. Since the crack moves very fast the stress in the strip will remain at
this large value even when the crack tip has moves far away from the strip as in position B. However, due to viscoelastic relaxation
the stress will finally arrive at the relaxed value $\sigma=E_0 \epsilon$ as at position C. 
The time this takes depends on the nature of the viscoelastic
relaxation process, e.g., for a process characterized by a single relaxation time $\tau$, 
a time $t> \tau$ (and distance $s>v\tau$) would be needed
to reach the relaxed state. During this relaxation mechanical energy is converted into heat. Since the crack tip is far away from
the region where this relaxation process takes place, it does not know about it, and the interfacial binding energy is converted into 
elastic energy in the rapid stretching of the strip in the process going from strip position A to B. Thus
$G_0 \Delta x = E_1 \epsilon_0^2 h_0 \Delta x/2$. However, the crack propagation energy $G$ refer to the relaxed state configuration
so that $G \Delta x = E_0 \epsilon_0^2 h_0 \Delta x/2$. Thus $G = E_0 \epsilon_0^2 h_0/2 = (E_0/E_1) E_1 \epsilon_0^2 h_0/2 = (E_0/E_1)G_0$.

This result for a closing crack tip can also be derived using the same approach as used for the opening crack. Thus in the present case,
for a fast moving closing crack the strain rather then the stress is known: $\epsilon(t)=0$ for $t<0$ and $\epsilon(t)=\epsilon_0$ for $t>0$.
Thus $\dot \epsilon (t) = \epsilon_0 \delta (t)$ and
$$U=-\int_{-\infty}^\infty dt \ \sigma \dot \epsilon = -\epsilon_0 \sigma (0) =  -\epsilon_0 {1\over 2} \left (\sigma(0^+)+\sigma(0^-) \right )$$
Since $\sigma(0^-) = 0$ and $\sigma(0^+) = E_1 \epsilon_0$ we get
$$U=-{1\over 2} {E_1 \epsilon_0^2}$$
and the condition $U h_0 \Delta x + G_0 \Delta x=0$ gives
$${1\over 2} h_0 {E_1 \epsilon_0^2} = {1\over 2} h_0 {\sigma_0^2 \over E_0} {E_1\over E_0} = G_0$$
or $G=G_0 E_0/E_1$.

\vskip 0.2cm
{\bf D. Opening crack}

The discussion in Sec. C can be easily generalized to a crack moving at a finite speed in a
viscoelastic solid. As a strip $\Delta x$ of material moves through the crack tip region,
for an opening crack we assume it experience the stress 
$$\sigma (t) = \sigma_0 \ \ \ {\rm for}  \ \ \ t<-\tau^*$$
$$\sigma (t) = \sigma_0 {\tau^* -t \over 2 \tau^*} \ \ \ {\rm for}  \ \ \ -\tau^* <t<\tau^*$$
$$\sigma (t) = 0 \ \ \ {\rm for}  \ \ \ t> \tau^*$$
where $v \tau^* = a$ is the width of the crack tip process zone.
We get
$$\sigma (\omega) = {1\over 2 \pi} \int_{-\infty}^\infty dt \ \sigma (t) e^{i\omega t} = {i \tau^* \sigma_0 \over 2 \pi} 
{{\rm sin} \xi \over (i0^+ - \xi )\xi}\eqno(12)$$
where $\xi = \omega \tau^*$ and where $0^+$ is an infinitesimal positive number.
Substituting (12) in (10) and using $\sigma_0 = E_0 \epsilon_0$ gives
$$U_{\rm open} = (\epsilon_0 E_0 )^2 {1\over \pi} \int_0^\infty d\omega \ R(\omega) {1\over \omega} {\rm Im} 
{1\over E(\omega)}\eqno(13)$$
where
$$R(\omega) = \left ({{\rm sin} (\omega \tau^*)\over \omega \tau^*}\right )^2\eqno(14)$$
We expect $v \tau^* \approx a$, where $a$ is the crack tip radius. In Ref. \cite{Brener} we used a different approach where
$R(\omega)$ was replaced by 
$$F(\omega)=\left [1-\left ({ \omega \over \omega_{\rm c}}\right )^2\right ]^{1/2}$$
where $\omega_{\rm c} = 2\pi v/a$. Note that if we expand $R$ and $F$ to quadratic order in $\omega$ then the two expressions agree if we choose
$\tau^* = (3/2)^{1/2} a/(2 \pi v)$.   

Energy conservation gives
$${1\over 2} h_0 \Delta x E_0 \epsilon_0^2 = h_0 \Delta x U_{\rm open}+ \gamma \Delta x$$
or using (13),
$${1\over 2} h_0 E_0 \epsilon_0^2 \left (1- E_0  {2\over \pi} \int_0^\infty d\omega \ R(\omega) {1\over \omega} {\rm Im} 
{1\over E(\omega)} \right )=\gamma$$
Since $\gamma = G_0$ and $G=h_0 E_0 \epsilon_0^2/2 $
we get
$$G={G_0 \over 
1- E_0  {2\over \pi} \int_0^\infty d\omega \ R(\omega) {1\over \omega} {\rm Im} 
{1\over E(\omega)}}\eqno(15)$$
Note that when $v\rightarrow \infty$ we have $\tau^* \rightarrow 0$ and hence $R\rightarrow 1$. Thus for very high
opening crack speeds
$$G (v=\infty)={G_0 \over 
1- E_0  {2\over \pi} \int_0^\infty d\omega \ {1\over \omega} {\rm Im} 
{1\over E(\omega)}}$$
Using (5) this gives $G (v=\infty)=(E_1/E_0) G_0$. 

Using (5) we can write (15) as
$${G_0 \over G} =1- {E_1  {2\over \pi} \int_0^\infty d\omega \ R(\omega) {1\over \omega} {\rm Im} 
{1\over E(\omega)} \over 1+ E_1  {2\over \pi} \int_0^\infty d\omega \  {1\over \omega} {\rm Im} 
{1\over E(\omega)}}\eqno(16)$$
which is convenient for numerical calculations.

Note that $G$ depends on the crack tip size parameter $a$. Experiments have shown that the
crack tip radius increases with the crack tip speed. We can choose $a$ so that the stress
for $r=a$ is of order characteristic yield stress $\sigma_{\rm c}$, e.g., the stress to break
bonds, which could be strong covalent bonds for cohesive crack propagation. The stress close to the crack tip
is given by
$$\sigma \approx {C \over r^{1/2}}$$
At a distance $\sim h_0$ from the crack tip the stress is of order $\sigma_0$ so we expect $C / (\alpha h_0)^{1/2} \approx \sigma_0$,
where $\alpha$ is a number of order unity.
Thus the stress at the crack tip $r=a$ is $\sigma = \sigma_{\rm c} \approx \sigma_0 (\alpha h_0/a)^{1/2}$,
or using  $G= h_0 E_0 \epsilon_0^2/2 = h_0 \sigma_0^2/(2E_0)$ we get
$$a = {E_0 2 \alpha G\over  \sigma_{\rm c}^2}.\eqno(17)$$
If we choose $\alpha = 1/(4 \pi)$ we obtain the equation derived in Ref. \cite{Brener}.
The tip radius $a(v)$ depend on the crack tip speed $v$, and using that $G(\infty)=G_0 E_1/E_0$
we get $a(\infty) = a_0 E_1/E_0$ where $a_0 = a(0)$ is the crack tip radius for very low crack tip speed.

\vskip 0.2cm
{\bf E. Closing crack}

The strain
$$\epsilon (t) = 0 \ \ \ {\rm for}  \ \ \ t<-\tau^*$$
$$\epsilon (t) = \epsilon_0 {t+\tau^* \over 2 \tau^*} \ \ \ {\rm for}  \ \ \ -\tau^* <t<\tau^*$$
$$\epsilon (t) = \epsilon_0 \ \ \ {\rm for}  \ \ \ t> \tau^*$$
We get
$$\epsilon (\omega) = {1\over 2 \pi} \int_{-\infty}^\infty dt \ \epsilon (t) e^{i\omega t} = {i \tau^* \epsilon_0 \over 2 \pi} 
{{\rm sin} \xi \over (i0^+ +\xi)\xi}\eqno(18)$$
where $\xi = \omega \tau^*$ and where $0^+$ is an infinitesimal positive number.
Substituting (18) in (9) gives
$$U_{\rm close} = \epsilon_0^2 {1\over \pi} \int_0^\infty d\omega \ R(\omega) {1\over \omega} {\rm Im} [-E(\omega)]\eqno(19)$$
For closing cracks the energy conservation condition gives
$$\gamma \Delta x = {1\over 2} h_0 \Delta x E_0 \epsilon_0^2 + h_0 \Delta x U_{\rm close}$$
or using (19),
$$\gamma  = {1\over 2} h_0  E_0 \epsilon_0^2 \left (1+E_0^{-1} {2\over \pi} \int_0^\infty 
d\omega \ R(\omega) {1\over \omega} {\rm Im} [-E(\omega)]\right )\eqno(20)$$
or
$$G={G_0 \over  
1+E_0^{-1} {2\over \pi} \int_0^\infty d\omega \ R(\omega) {1\over \omega} {\rm Im} [-E(\omega)]}$$
Note that when $v\rightarrow \infty$ we have $\tau^* \rightarrow 0$ and hence $R\rightarrow 1$. Thus for very high
opening crack speeds
$$G(v=\infty)={G_0 \over  
1+E_0^{-1} {2\over \pi} \int_0^\infty d\omega \ {1\over \omega} {\rm Im} [-E(\omega)]}$$
Using (6) this gives $G (v=\infty)=(E_0/E_1) G_0$. Using (6) we can write (20) as
$${G_0 \over G} =  
1+{E_1^{-1} {2\over \pi} \int_0^\infty d\omega \ R(\omega) {1\over \omega} {\rm Im} [-E(\omega)] \over 1- 
E_1^{-1} {2\over \pi} \int_0^\infty d\omega \ {1\over \omega} {\rm Im} [-E(\omega)] }\eqno(21)$$

For the crack opening we determined the crack tip radius $a$ such as the stress at the crack 
tip equal the critical stress necessary for bond breaking.
This resulted in a radius which, in agreement with experiments, increases with increasing crack tip speed
and in particular $a\rightarrow a_0 E_1/E_0$ as the crack tip velocity $v \rightarrow \infty$. 
However, making the same assumption for the closing crack result in unphysical results,
namely $a\rightarrow a_0 E_0/E_1$. We expect $a_0$ to be of order 1 nm so in a typical case 
with $E_0/E_1 \approx 10^{-3}$ the crack tip radius $a\rightarrow 0.001 \times a_0= 0.01 \ {\rm nm}$.
But this result is unphysical; the radius cannot be smaller than an atomic length and in fact we 
expect $a\approx a_0$ for all velocities for a closing crack.

Now, if we choose $a=a_0$ for all crack tip velocities for the closing crack, then the 
stress $\sigma_1$ at the crack tip for high crack-tip speed would be much smaller than the adhesive bonding stress
at the crack tip. This imply that large forces will act on the rubber segments at the crack tip 
and the rubber segments will accelerate and snap into contact, and perhaps
undergoes some other rapid event where energy is lost (converted into heat). 
In fact, Carbone et al have suggested that some slip will occur in the crack tip process zone
during contact formation for soft adhesive films. This will 
make $G_0$ smaller than in the adiabatic limit (since for a closing 
crack $G_0(v)=\gamma-w$ is the binding energy per unit surface energy, $\gamma$,
minus the energy $w$ dissipated in the crack tip process zone).
The combination (for high crack tip speed) of the viscoelastic reduction factor
$E_0/E_1$, and the reduction in $G_0$ with increasing crack tip speed 
will make adhesion nearly absent during contact formation in typical cases. 
This is in accordance with experiments where for macroscopic solids, even for solids with very smooth surfaces, 
in most cases no adhesion can be observed during contact formation (see Sec. 4).

Here we note that the problem addressed above, involving how to determine the crack tip radius $a$, and the related
stress mismatch problem, also occur in a modified form in the Barenblatt process zone treatment of the closing crack problem.
Thus for a fast moving crack a region of compressible stress occur close to the crack tip for which no 
physical explanation exist\cite{HuiX,Green1}.

\begin{figure}[tbp]
\includegraphics[width=0.45\textwidth,angle=0]{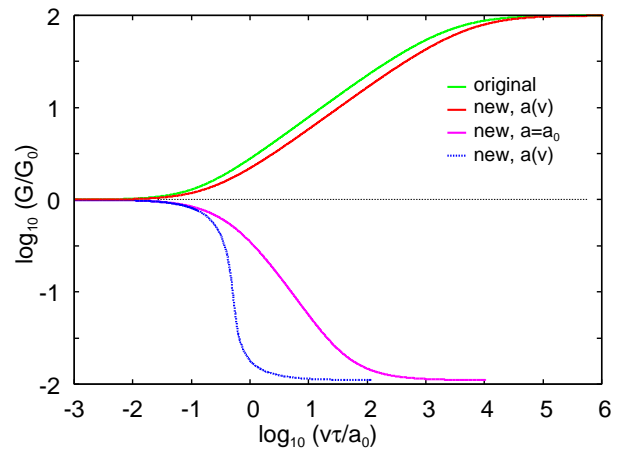}
\caption{
The crack propagation energy as a function of the crack tip speed (log-log scale) using the rheological
model defined by (1).
The green line is for an opening crack using the original Persson-Brener theory (Ref. \cite{Brener}) 
where the viscoelastic energy dissipation is calculated from the crack tip stress field. 
This result is virtually identical to the result obtained using the Barenblatt process cone model\cite{Green1}. 
The red line is the result for an opening crack using the simplified treatment, where the viscoelastic energy dissipation
is estimated (from (16) and (17)) using the stretching-segment model (see Sec. 3). 
The dashed blue line is the (unphysical) result obtained from (21) and (17), where the crack tip radius
becomes unphysical small for high crack tip speed ($a\rightarrow (E_0/E_1)a_0$ as $v\rightarrow \infty$). The
pink line is the result for closing crack using (21) and assuming a constant crack tip radius $a=a_0$.
}
\label{1logv.2logG.3.cases.eps}
\end{figure}

\vskip 0.2cm
{\bf 3 Numerical results}

We have calculated the crack propagation energy $G(v)$ using the simple three-element rheology model
shown in Fig. \ref{rheologypic.eps}.
In Fig. \ref{1logv.2logG.3.cases.eps} we show
the crack propagation energy as a function of the crack tip speed (log-log scale) using the rheological
model defined by (1).
The green line is for an opening crack using the original theory (Ref. \cite{Brener}) 
where the viscoelastic energy dissipation is calculated from the crack tip stress field. 
This result is virtually identical to the result obtained using the Barenblatt process cone model\cite{Green1}. 
The red line is the result for an opening crack using the simplified treatment where the viscoelastic energy dissipation
is estimated (from (16) and (17)) using the stretching-segment picture. 
The dashed blue line is the (unphysical) result obtained from (21) and (17), where the crick tip radius
becomes unphysical small for high crack tip speed ($a\rightarrow (E_0/E_1)a_0$ as $v\rightarrow \infty$). The
pink line is the result for closing crack using (21) and assuming a constant crack tip radius $a=a_0$.
Comparing the pink line with the red (and green) lines we conclude that if we write for crack opening
$G=G_0 (1+f(v))$ then for small $v\tau /a_0$ we have for crack closing $G \approx G_0 /(1+f(v))$. In the
Barenblatt crack zone treatement this relation is found to hold approximately for all crack tip velocities\cite{Green1}.

\begin{figure}[tbp]
\includegraphics[width=0.45\textwidth,angle=0]{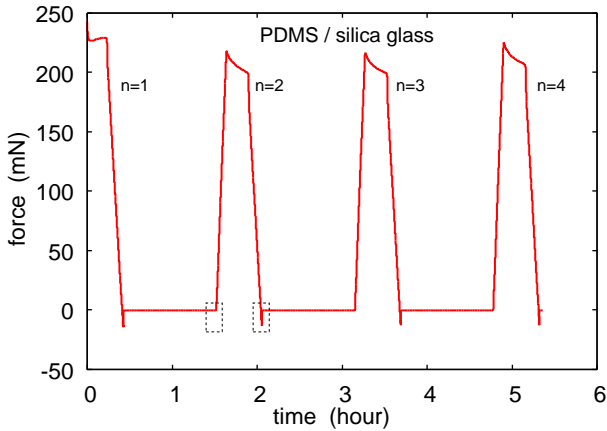}
\caption{
The interaction force between a glass ball (diameter $2R=4 \ {\rm cm}$) moved in repeated contact with a 
flat PDMS surface. The approach and retraction speed is $v_z=0.33  \ {\rm \mu m/s}$. The dashed rectangular regions
are shown magnified in Fig. \ref{1time.Force.PDMS.B.eps} (crack opening) 
and Fig. \ref{1time.Force.PDMS.C.eps} (closing crack). 
}
\label{1time.Force.PDMS.A.eps}
\end{figure}

\begin{figure}[tbp]
\includegraphics[width=0.45\textwidth,angle=0]{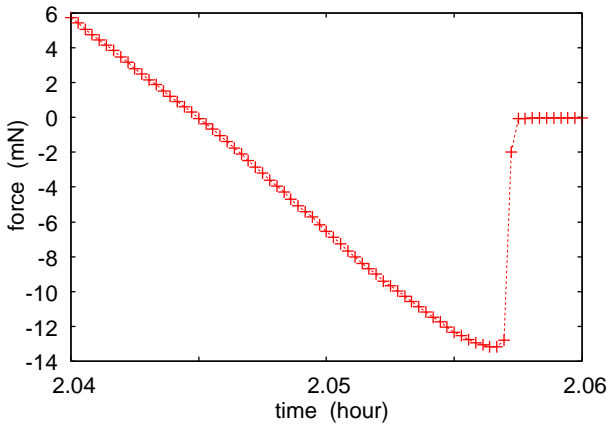}
\caption{
Magnified view of the second pull-off event in Fig. \ref{1time.Force.PDMS.A.eps}. The
(radial) crack tip speed just before snap-off is $v_r \approx 14 \ {\rm \mu m/s}$.
The pull-off force correspond to the work of adhesion $G\approx 0.14 \ {\rm J/m^2}$.
There is one data point per second.
}
\label{1time.Force.PDMS.B.eps}
\end{figure}

\begin{figure}[tbp]
\includegraphics[width=0.45\textwidth,angle=0]{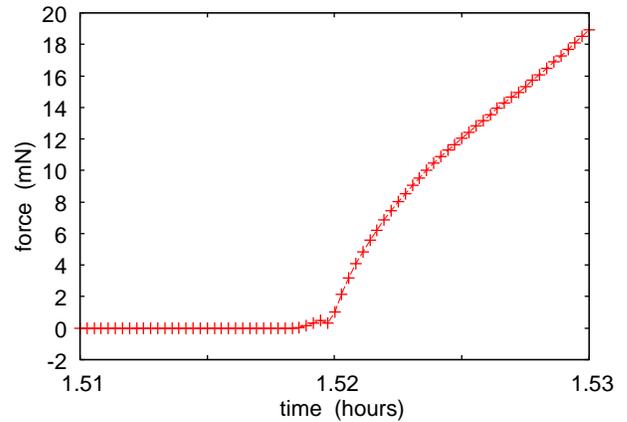}
\caption{
Magnified view of the second contact formation event in Fig. \ref{1time.Force.PDMS.A.eps}. 
Note the strong adhesion hysteresis: no adhesion is observed during approach but adhesion
is observed during pull-off (see Fig. \ref{1time.Force.PDMS.B.eps}).
There is one data point per second.
}
\label{1time.Force.PDMS.C.eps}
\end{figure}

\begin{figure}[tbp]
\includegraphics[width=0.45\textwidth,angle=0]{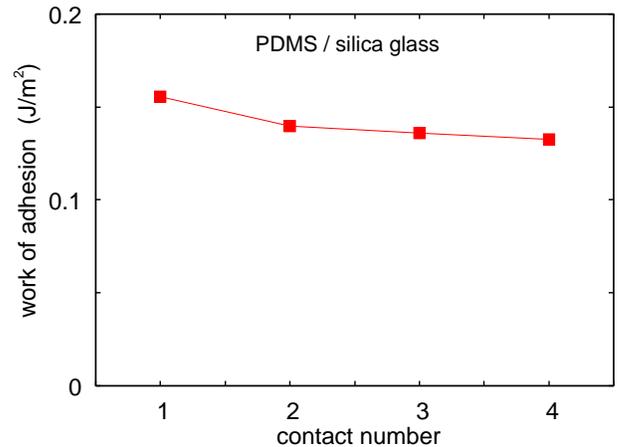}
\caption{
The work of adhesion during pull-off as a function of the contact number.
The decrease in the work of adhesion is due to transfer of molecules from the
PDMS to the glass ball, passivating the glass surface.
}
\label{1number.2WorkAdhesion.eps}
\end{figure}

\begin{figure}[tbp]
\includegraphics[width=0.45\textwidth,angle=0]{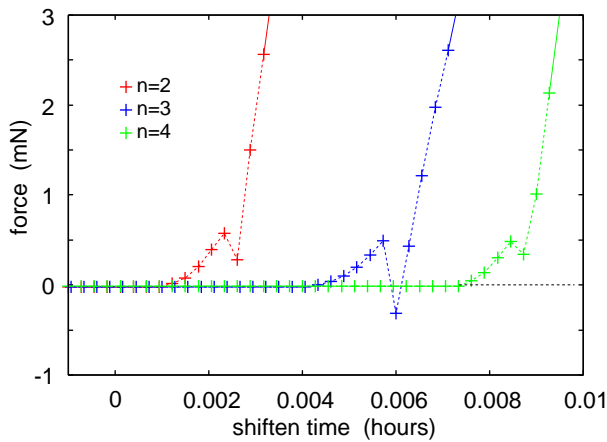}
\caption{
The interaction force between the glass ball and the PDMS surface on approach
for the contacts $n=2$, 3 and 4 in Fig. \ref{1time.Force.PDMS.A.eps}.
Note the small repulsive barrier before contact which 
we attribute to the influence of dust particles in the nominal contact region.}
\label{1time.2Force.PDMS.approach1.eps}
\end{figure}

\vskip 0.2cm
{\bf 4 Discussion}

For a very slowly moving adhesive crack the elastic stress at the crack tip must be just balancing the adhesive stress
so no rapid non-thermal instabilities, such as snap-off or snap-in, can occur. 
We have argued above that for a fast moving closing adhesive crack the crack propagation energy $G(v)$ must be reduced
not only by the viscoelastic factor $E_0/E_1$ but also $G_0(v)$ must be strongly reduced 
due to rapid events in the crack-tip process zone caused by the stress mismatch at the crack tip.
This conclusion is supported by adhesion experiments. Thus, adhesion is usually not observed
when two macroscopic solids approach each other, while for elastically soft solids
strong adhesion may be observed upon separation. As an example, in Fig. 
\ref{1time.Force.PDMS.A.eps} we show the interaction force between a glass ball (radius $R=2 \ {\rm cm}$) and a
flat PDMS rubber surface, both with very smooth surfaces. The ball moves up and down with the speed
$0.33 \ {\rm \mu m}$ and we show the interaction force for 4 contacts. Note that 
no attraction is detected during contact formation, but during pull-off adhesion manifest itself as a negative interaction force.
This is illustrated in detail (for the second contact cycle) in Fig. \ref{1time.Force.PDMS.B.eps} (pull-off) 
and  Fig. \ref{1time.Force.PDMS.C.eps} (contact formation).
The work of adhesion during pull-off as a function of the contact number is shown in Fig. \ref{1number.2WorkAdhesion.eps}.
The decrease in the work of adhesion with increasing number of contacts is due to transfer of molecules from the
PDMS to the glass ball, passivating the glass surface.

For the second pull-off the work of adhesion $G\approx 0.14 \ {\rm J/m^2}$. This is a factor of $\sim 2$ larger than
the adiabatic work of adhesion
between PDMS and a glass surface, which is about $G_0 \approx 0.06 \ {\rm J/m^2}$. This imply that the viscoelastic enhancement
factor $1+f(v,T)$ is a factor of $\approx 2$ (or less), in agreement with calculations\cite{softM}. 
Thus if viscoelasticity would be the only energy dissipation process
we would expect a work of adhesion during contact formation to be
$G_{\rm close} \approx G_0/(1+f(v,T)) \approx 0.03 \ {\rm J/m^2}$, corresponding to an attractive ball-PDMS force of
$\approx 3 \ {\rm mN}$. However, the small dip in the measured
contact formation force is 
between $0.1-0.8 \ {\rm mN}$ (see Fig. \ref{1time.2Force.PDMS.approach1.eps}). 
This imply that some energy dissipation process (with the dissipated energy per unit surface area
$w$) must occur in the crack tip process zone during crack
closing. In this case part of the energy (per unit surface area) $\gamma$ gained in the bond formation process
is lost in the crack tip process zone and $G_0 = \gamma - w < \gamma$. 
The exact processes occurring is not known but may involve some snap-in or 
local slip at the contacting interface. 

Before the small adhesive dip in the time-force curves in Fig. \ref{1time.2Force.PDMS.approach1.eps} 
the ball-flat interaction is repulsive. There are at least two possible origins of this repulsion.
One effect is squeeze-film:
When the ball is very close to the PDMS surface a hydrodynamic pressure builds up in 
the air film between the ball and the flat
PDMS surface. This force can be estimated using the Navier Stokes equations of fluid dynamics 
(on the simplified Reynolds equation form).
Thus, if $h(t)$ denote the shortest ball-flat separation, then for $h<<R$ (see Ref. \cite{squeeze})
$$F = 6 \pi \mu R^2 {\dot h \over h}$$
This equation gives a similar dependency on the separation $h$ as shown in Fig.
\ref{1time.2Force.PDMS.approach1.eps}, but using the viscosity of air
($\mu \approx 1.8\times 10^{-5} \ {\rm Pas}$) the magnitude of the calculated force is a factor 
of $\sim 1000$ too small. Another explanation is that there are one or several
dust particles adsorbed on the PDMS surface, which need to be squeezed into the rubber surface (elastic deformation) 
before the glass-PDMS contact can occur. Since the experiments was performed in the normal atmosphere this is a likely explanation.

When an opening crack propagate in the bulk of an elastomer (cohesive crack) strong covalent bonds are broken at the crack tip.
In a recent study\cite{PRX} using fluorogenic mechanochemistry with
quantitative  confocal  microscopy  mapping,  it was found how  many  and  where  covalent
bonds are broken. The measurements reveal that  bond scission
near the crack plane can be delocalized over up to hundreds of micrometers and increase
$G_0$ by  a  factor  of  $\approx 100$  depending  on  temperature  and  stretch  rate,  pointing  to  an  intricated
coupling   between   strain   rate   dependent   viscous   dissipation   and   strain   dependent
irreversible  net
work  scission.  These  findings shows that energy dissipated by covalent bond scission accounts for a much larger
fraction  of  the  total  fracture  energy  than  previously  believed.

The study above does not give any dependency of the crack propagation energy on
the height $h_0$ of the viscoelastic slab. That is, the crack propagation factor $G/G_0 = 1+f(v,T)$
does not depend on the height $h_0$ of the rubber sample assuming the length (in the $x$-direction) 
$L$ of the sample is infinite. This differ from the conclusion derived in Ref. \cite{small}
and the proposal by de Gennes\cite{Gennes} that the origin of instabilities in the pull-off of adhesive tape
may be due the influence of the finite film thickness on the viscoelastic energy dissipation. 
However, the present study does not include the singular part of 
the crack tip stress field, and more studies are needed to understand the role of finite-size effects
on the viscoelastic contribution to the crack propagation energy. 

\vskip 0.2cm
{\bf 5 Summary and conclusion}

I have studied crack propagation in a stretched rectangular strip of a viscoelastic solid.
I have shown that for an opening crack using a very simple model for the stress field, which describe the stress
correctly far from the crack tip but not close to it, gives a viscoelastic crack propagation energy factor
$G/G_0=1+f(v,T)$ very close to the one obtained using the Barenblatt process zone model, or the
Persson-Brener crack tip model. For a closing crack tip the same approach gives physical reasonable
result only if one assume a constant crack-tip radius and assume that the crack tip process zone
energy $G_0$ decreases with increasing crack tip speed, which imply that for high enough crack tip speed
dissipate processes, e.g. involving rapid flipping of molecular segment or local slip,
occur at the crack tip during closing, which consumes a large fraction of the gain in energy due to
the bond formation.

\end{document}